# A new "111" type iron pnictide superconductor LiFeP

Z. Deng, X. C. Wang, Q.Q. Liu, S. J. Zhang, Y. X. Lv, J. L. Zhu, R.C. Yu, C.Q. Jin

*Institute of Physics, Chinese Academy of Sciences, Beijing, China*

## Abstract

A new iron pnictide LiFeP superconductor was found. The compound crystallizes into a $Cu_2Sb$ structure containing an "FeP" layer showing superconductivity with maximum $T_c$ of 6K. This is the first "111" type iron pnictide superconductor containing no arsenic. The new superconductor is featured with itinerant behavior at normal state that could be helpful to understand the novel superconducting mechanism of iron pnictide compounds.

# Introduction

The discovery of iron pnictide superconductors [1,2] opens a new era for unconventional superconductivity. Similar to high Tc cuprates, the superconductivity in iron arsenide compounds is related to a layered structure[1].The superconducting compounds consist of an iron pinictide layer that supports the superconducting current interlaced with a charge reservoir layer. The charge reservoir layer can be rare earth oxide or fluoride/alkaline earth fluoride, alkaline earth metal or alkaline metal that gives rise to a "1111" system [1~6], "122" system[7] or "111" system [8], respectively. But unlike the cuprates, which are strongly correlated charge transfer type Mott compounds, the layered iron arsenide is an itinerant metal/semi-metal. The high transition temperature in this itinerant system containing the magnetic element Fe challenges the conventional BCS mechanism. Therefore fundamentally it is very interesting to question the underlying superconducting mechanism of the iron pnictide superconductors. Seaching for new superconductors of simpler structure will be helpful in the study of the physical mechanism of the unusual superconductivity in iron pnictide compounds, particularly those containing pnictide element other than As. Here we report that iron phosphide LiFeP was found to be a new superconductor. The compound of LiFeP crystallizes into a structure similar to LiFeAs[8] as shown in Fig. 1. The compound becomes superconductive with Tc up to 6K.

# **Experimental**

The LiFeP compounds are synthesized using solid state reaction method. The starting materials of Li (99.9%) and Fe are mixed according to the nominal formula LiFeP.

The Fe precursors are synthesized from high purity Fe and P powders that are sealed into an evacuated quartz tube. The mixtures are sintered at 800°C for 10 h. Since the compositions of LiFeP are either hygroscopic or easy to react with oxygen or nitrogen, all the process is performed in the protection of high purity Ar inside a glove box. We found that pressure can assist the synthesis of "111" type LiFeP since high pressure can effectively prevent lithium from oxidizing or evaporating upon heating. The pellets of mixed starting materials wrapped with gold foil are sintered at 1.8GPa, 800°C for 60 min. Alternatively this sample can be synthesized using traditional methods by sealing the components into a quartz tube. The quartz tube is sintered at 800°C for 30 h. The recovered samples are characterized by x-ray powder diffraction with a Philips X'pert diffractometer using $CuK_{\alpha 1}$ radiation. Diffraction data were collected with 0.02° and 15 s /step. Rietveld analysis has been performed by using the GSAS program software package . The DC magnetic susceptibility was characterized using SQUID magnetometer (Quantum design) while the electric conductivity as well as the specific heat was measured using the standard four probe method with a PPMS system.

## Results and Discussion

Figure 1 shows the X-ray diffraction patterns of a LiFeP sample. The phase can be indexed quite well into a tetragonal structure [9] of space group P4/nmm. Figure 1 also shows a schematic view of the crystal structure of LiFeP. The lattice parameters obtained for LiFeP are $a$=3.692Å, $c$=6.031 Å. Compared with "111"LiFeAs where

$a$=3.771Å, $c$=6.357Å the *ab* plane is shrunk by 2.1% for "111" LiFeP. This is comparable to the 1.8% change of lattice parameters for "1111" type LaFePO[1] with $a$ =3.964Å and $c$ =8.512 Å versus "1111" type LaFeAsO with $a$ =4.036Å and $c$ =8.739 Å [2]. The diffraction pattern was refined using the Rietveld method. The results are listed in Table 1 where the [FeP$_4$] tetrahedron coordination is highlighted in the inset of Fig.1(b). It was observed from experiments that the angle between pnictide ~iron ~pnictide bond is correlated with the superconducting transition temperature [10,11]. The first principle calculations indicated that the Fermi surface (FS) of iron pnictide superconductor consists of three hole like pockets centered at Γ point plus two electron like parts at M point [12]. The topology of FS is very sensitive to the crystal geometry [12,13] such as to the iron pnictide polyhedron. It is worth of mention that the [FeP$_4$] tetrahedron of LiFeP is only slightly distorted from the ideal case of $\alpha=\beta=108°$ with $\alpha$=105.58° & $\beta$=109.92°. This is in sharp contrast either with "1111" type iron phosphide LaFePO[1] where $\alpha$=120.18° & $\beta$=104.39°, or with "1111" type iron arsenide LaOFeAs[2],"122" AFe2As2[7] or "111" type LiFeAs [8,14,15]. The"111" type LiFeP is the less distorted in terms of iron pnictide tetrahedron that makes it a model to study the property of an ideal iron pnictide coordination. Figure 2 shows the temperature dependence of the electric conductivity of LiFeP. Although we change the nominal composition of Li content from 0.8 to 1.2, samples show almost the same superconducting transition with Tc$^{onset}$ ~ 6K. This is quite similar to "111" type LiFeAs where the superconducting transition temperature seems not sensitive to the nominal Li content [8], but much different from "1111"

type[1~6] or "122" type [7] iron pnictide system. Especially the "1111" phosphide LaFePO shows a Tc change with doping [1]. Here the "111" pnictide system presents the same inert reaction to "chemical doping" for either LiFeAs or the present LiFeP. The normal state of LiFeP shows good metallic behavior without an abrupt change of resistivity that usually proceeds the spin density wave (SDW) as manifested for "1111" [2~6] or "122"[7] type iron arsennide system. The magnetic ordering of iron arsenide system was well recorded with neutron diffraction to support the SDW state[16]. However the SDW related resistance drop is absent in "1111" type LaFePO or "111" type LiFeAs. The first principle calculations for LiFeAs indicated little change of density of states (DOS) or Fermi surface topology with Li content that could account for its small influence on Tc[17]. The same situation may be valid for LiFeP case. Figure 2(b) shows the resistance versus temperature curve for LiFeP with H = 0, 0.1 & 1T. There is a very slight magnetic resistance effect at the normal state. Figure 3(a) shows the DC magnetic susceptibility of a LiFeP sample measured in both ZFC and FC mode with H=10 Oe. The large Meissner signal indicates the bulk superconducting nature of the sample. This is verified with specific heat measurement as shown in Figure 3(b) where a jump at Tc clearly indicates the bulk superconducting nature of the sample. Fitting the low temperature specific data above Tc using the formula $C=\gamma T+\beta_3 T^3$ gives rise to the electronic specific coefficient $\gamma = 16.08$ mJmol$^{-1}$K$^{-2}$. The result suggests the mass enhancement of carriers due to electron correlation. But comparing with high Tc cuprates the iron phosphide is more itinerant as suggested in recent ARPES studies on LaFePO[18].

In summary LiFeP was found to be a new iron pnictide superconductor. The compounds form a layered [FeP] structure similar to "111" LiFeAs. The superconductivity with Tc up to 6K was achieved.

**ACKNOWLEDGMENT**: This work was supported by NSF & MOST of China through research projects.

References:

1. Y. Kamihara, H. Hiramatsu, M. Hirano, R. Kawamura, H. Yanagi, T. Kamiya, and H. Hosono, **J. Am. Chem. Soc.128**, 10012(2006)

2. Y. Kamihara, T. Watanabe, M. Hirano, and H. Hosono, **J. Am. Chem. Soc.130**, 3296(2008).

3. X. H. Chen, T. Wu, G. Wu, R. H. Liu, H. Chen, D. F. Fang, **Nature 453**, 761(2008)

4. G. F. Chen, Z. Li, D. Wu, G. Li, W. Z. Hu, J. Dong, P. Zheng, J. L. Luo, and N. L. Wang, **Phys. Rev. Lett. 100**, 247002(2008

5. Z.-A Ren, J. Yang, W. Lu, W. Yi, X.-L Shen, Z.-G Li, G.-C Che, X.-L Dong, L.-L Sun, F. Zhou, Z.-X Zhao, **Europhys. Lett. 82**, 57002(2008)

6. H. H. Wen, G. Mu, L. Fang, H. Yang, X. Y. Zhu, **Eur. Phys. Lett. 82**, 17009(2008)

7 M. Rotter, M. Tegel, and D. Johrendt, **Phys. Rev. Lett. 101**, 107006 (2008)

8. X.C.Wang, Q.Q. Liu, Y.X. Lv, W.B. Gao, L.X.Yang, R.C. Yu, F.Y.Li, C.Q. Jin, **Solid State Communications 148,** 538(2008)

9 Von. R. Juza & K. Langer, **Z. Anorg. Allg. Chem 361**, 58(1968)

10 N. Takeshita, A. Iyo, H. Eisaki, H. Kito, and T. Ito, **J. Phys. Soc. Jap 77**, 75003(2008).

11 J.G. Zhao, L.H. Wang, D.W. Dong, Z.G. Liu, H.Z. Liu, G.F. Chen, D. Wu, J.L. Luo, N.L .Wang, Y. Yu, C. Q. Jin, Q.Z. Guo, **Journal of the American Chemical Society 130**, 13828 (2008)

12 S.Lebegue, **Phys. Rev. B 75**, 035110 (2007).

13 X.Dai, & Z. Fang, Y. Zhou. F. C. Zhang, **Phys. Rev Lett 101**, 57008(2008).

14 Michael J. Pitcher, Dinah R. Parker, Paul Adamson, Sebastian J. C. Herkelrath, Andrew T. Boothroyd, Simon J. Clarke, **Chem. Commun.**, 5918(2008).

15 Joshua H. Tapp, Zhongjia Tang, Bing Lv, Kalyan Sasmal, Bernd Lorenz, Paul C.W. Chu, Arnold M. Guloy, **Phys Rev. B 78**, 060505(2008).

16 D. J. Singh, **Phys Rev. B 78**, 094511(2008)

17 Clarina de la Cruz, Q. Huang, J. W. Lynn, Jiying Li, W. Ratcliff, J. L. Zarestky, H. A. Mook, G. F. Chen, J. L. Luo, N. L. Wang, Pengcheng Dai, **Nature 453**, 899(2008)

18 D. H. Lu, M. Yi, S.-K. Mo, A. S. Erickson, J. Analytis, J.-H. Chu, D. J. Singh, Z. Hussain, T. H. Geballe, I. R. Fisher, Z.X.Shen, **Nature 455**, 81(2008)

# Table 1

The refinement results of the X-ray diffraction patterns of LiFeP.

| Atom | site | x    | y    | z        | occupancy |
|------|------|------|------|----------|-----------|
| Li   | 2c   | 0.25 | 0.25 | 0.651962 | 0.8       |
| Fe   | 2a   | 0    | 0    | 0        | 1         |
| P    | 2c   | 0.25 | 0.25 | 0.220055 | 1         |

Space group: P4/nmm. Unit-cell dimensions: $a$=3.69239(2)Å, $c$=6.03081(2)Å; $R_{wp}$=5.3%, $R_p$=2.5%. The refinement range of $2\theta$ is 10-135°. CuK$\alpha_1$ radiation was used.

# Figure Captions:

**Fig.1:** Schematic view of the crystal structure of LiFeP; the X-ray diffraction pattern of LiFeP, refined with the Rietveld method into the "111" structure.

**Fig.2:** (a) The temperature dependence of resistivity for LiFeP of variant nominal Li content showing superconducting transition up to 6K with the metallic normal state; (b) the resistance versus temperature curve of LiFeP at H=0, 0.1 & 1T.

**Fig.3**: (a) The DC susceptibility of LiFeP in both ZFC)andFCmode; (b) the specific heat measurement with a sharp jump at the superconducting transition temperature, indicating the bulk superconducting nature.

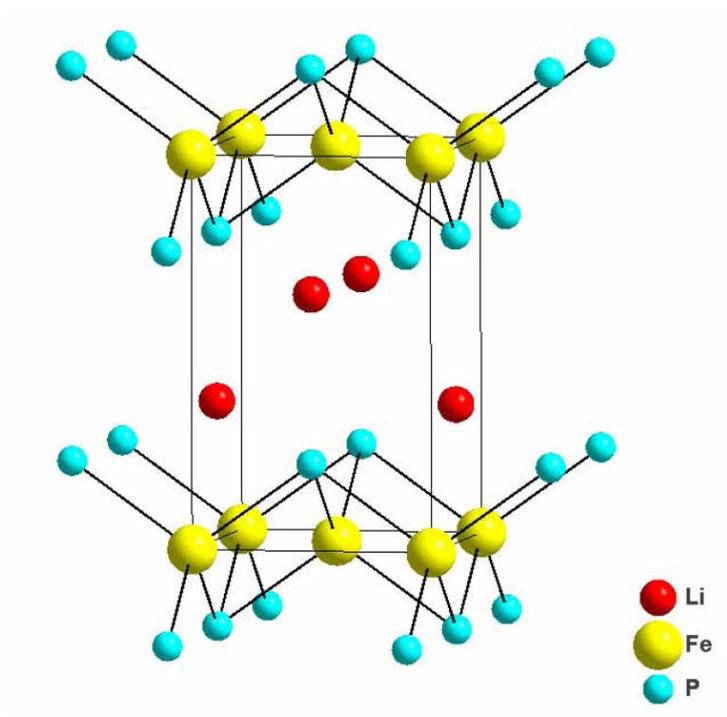

**Fig.1(a)**

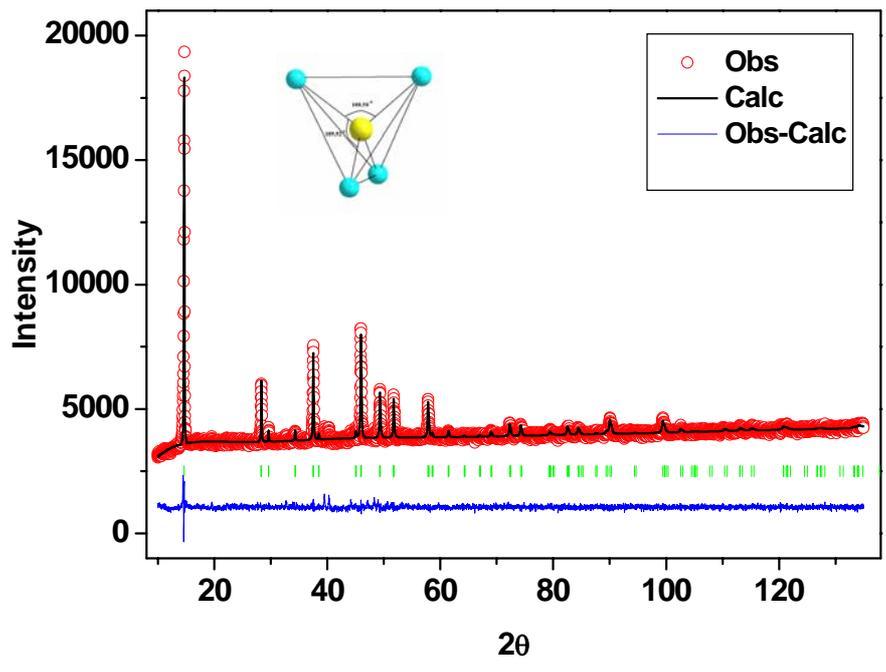

**Fig.1(b)**

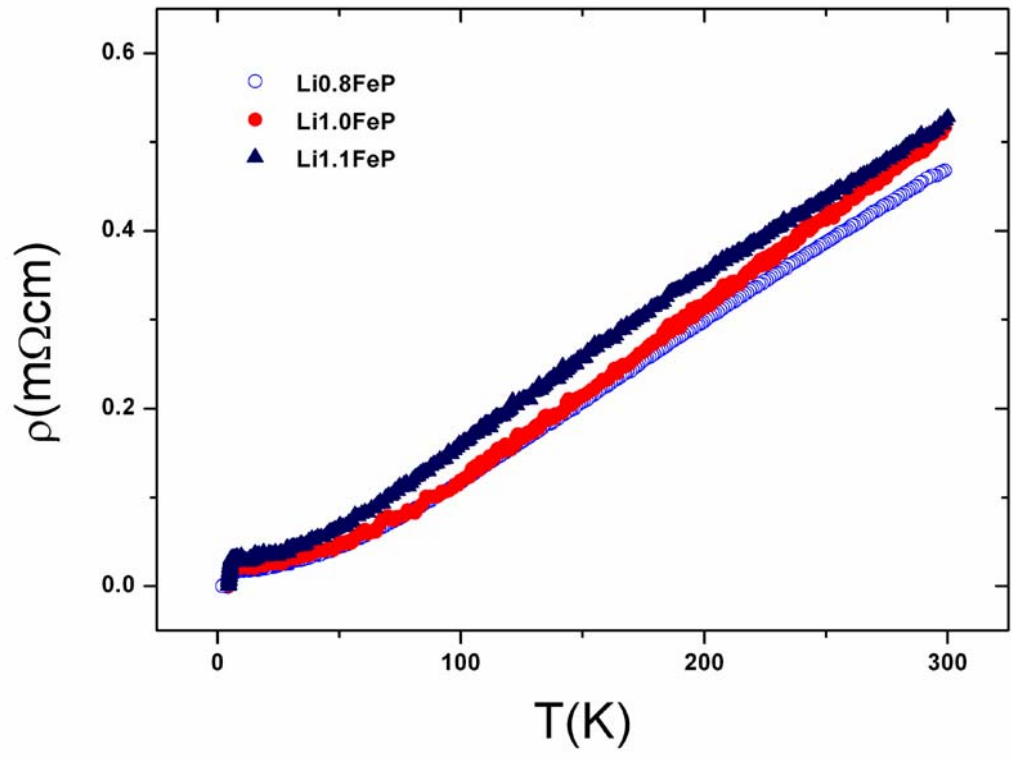

**Fig.2(a)**

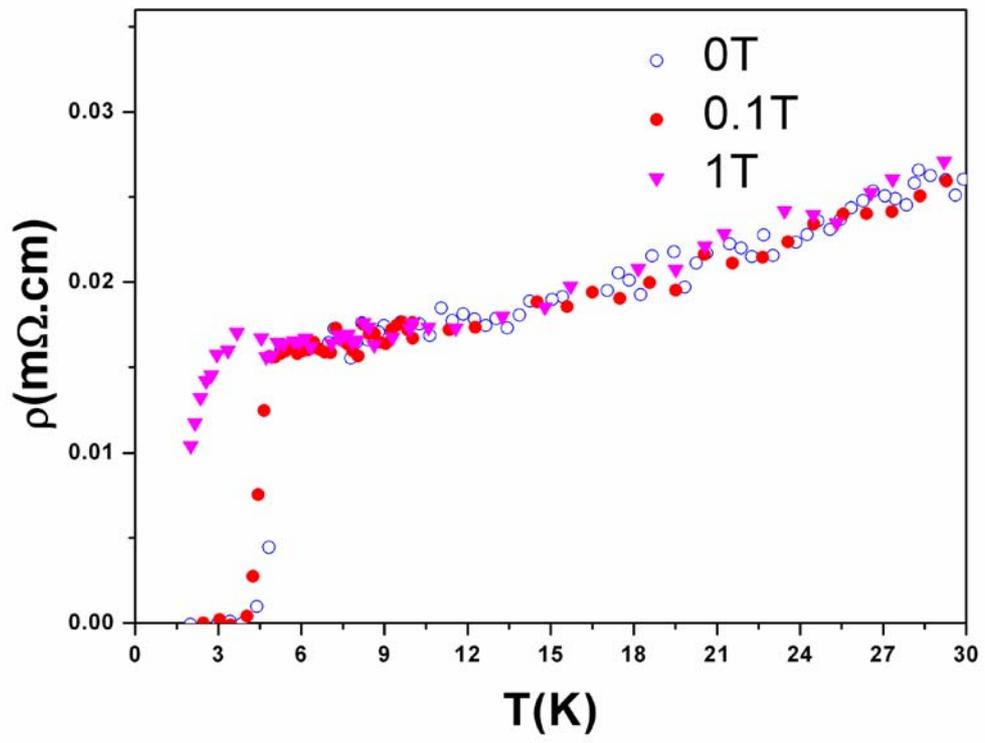

**Fig.2(b)**

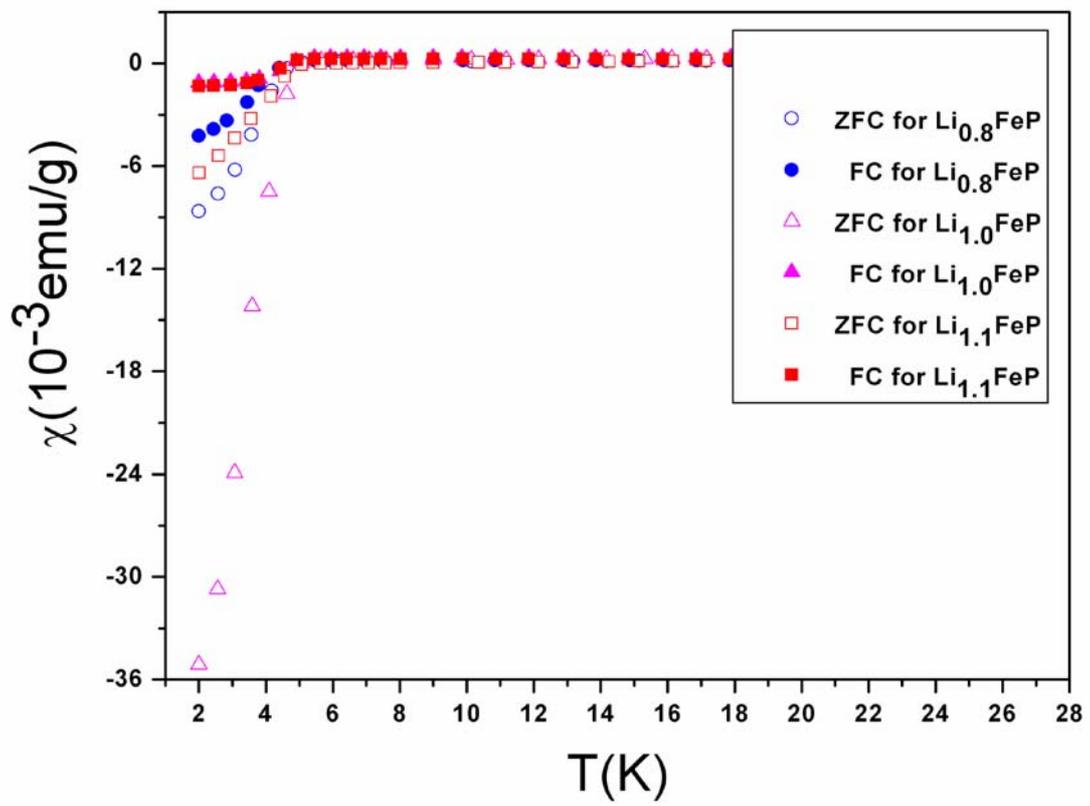

**Fig.3(a)**

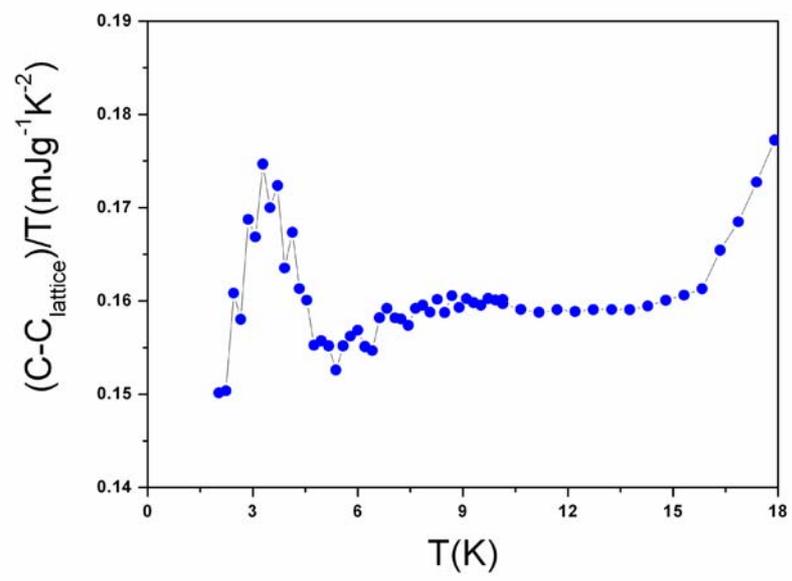

**Fig.3(b)**